\documentstyle[a4,epsf,cite,12pt]{article}
\newcommand{\be}{\begin{equation}}
\newcommand{\ba}{\begin{eqnarray}}
\newcommand{\ee}{\end{equation}}
\newcommand{\ea}{\end{eqnarray}}
\newcommand{\nn}{\nonumber}

\newcommand{\GeV}{\;\mbox{GeV}}
\newcommand{\eV}{\;\mbox{eV}}

\newcommand{\secs}{\;\mbox{s}}

\newcommand{\ol}{\overline}

\renewcommand{\i}{\mbox{\small i}}

\newcounter{currequation}
\begin{document}
\title{CP noninvariance and an effective cosmological constant: the energy
density in a pseudoscalar field which arises from a cosmological, spontaneously-broken
chiral symmetry}
\author{Saul Barshay and Georg Kreyerhoff\\
III.~Physikalisches Institut A\\
RWTH Aachen\\
D-52056 Aachen}
\maketitle
\begin{abstract}
We present and discuss the properties and the main results of a cosmological model
with a spontaneously-broken chiral symmetry. The model contains and relates dynamically,
two spin-zero fields. The scalar field can provide the dynamical basis
for inflation in the early universe. The pseudoscalar, Goldstone field can provide
an early, small residual vacuum energy density, the absolute value of which we estimate
to be similar to the present, empirically small vacuum energy density. The small 
energy scale for this effective cosmological constant is estimated separately, by relating it
dynamically to the empirical, small scale of neutrino mass. CP invariance is broken spontaneously.
This provides a natural basis for the early generation of an antineutrino-neutrino asymmetry, whose magnitude
we estimate, and find to be significant.
\end{abstract}
Speculations concerning the existence and the dynamics of fields with spin-zero and positive
parity play a central role in cosmological models \cite{ref1,ref2,ref3,ref4}. Two such (scalar)
fields are presently of particular interest in the construction of models for hypothetical
inflation in the early universe \cite{ref5}, and for a presumed small cosmological constant \cite{ref6,ref7}
in the present epoch \cite{ref3,ref8}. In these considerations, one issue that is always present (but is often
not explicitly stated), involves an effort to relate the existence of the hypothetical field to phenomena other
than the single one for which it is postulated, i.~e.~either to provide a large vacuum energy density
which drives early inflation \cite{ref5}, or to provide a small vacuum energy density which appears as an
effective cosmological constant today \cite{ref3}. One possibility involves considering the inflationary scalar field as the same
field whose energy density gives a present effective cosmological constant \cite{ref8}. A different
possibility involves considering the cosmological-constant scalar field as a field which gives mass
(varying with time) to dark-matter particles \cite{ref9,ref4}. These two
possibilities invoke (homogeneous) scalar field values which vary with increasing time, having large magnitudes of order 
of the Planck energy $M_P\cong 1.2\times 10^{19}\GeV$, in the present epoch. A further implication of the
assumed, unusual forms for the potential-energy density of the scalar field, is the existence of fluctuations
in the field which correspond to particles with miniscule effective mass, $\sim 10^{-32}\eV$ in the
present epoch (effective squared mass is the second derivative of the potential-energy density with respect
to the field) \cite{ref3}. Exchange of such scalar particles gives rise to very long-range,
{\underline{coherent}} forces \cite{ref3}. The potential-energy densities which are considered often involve
inverse powers (or inverse exponentials \cite{ref3}) of the scalar field $\phi$, (such as $M^8/\phi^4$)
\cite{ref8,ref9}, and consequently dimensional parameters (like $M$). The authors state \cite{ref9}
that the ``meaning within quantum field theory'' of such inverse powers of $\phi$, and the matter of the
``naturalness'' of dimensional parameters, is not presently considered. The justification for this is given as the fact
that the models are not meant to address the problem of the (possibly large) effective cosmological
constant which can arise from the energy scale of quantum fluctuations in the vacuum \cite{ref3}. Given 
the several curious aspects of models of the type mentioned above, it seems to us to be reasonable to
raise the following questions. Is it possible that a known type of model in quantum field theory which
inextricably contains two spin-zero fields, and which relates their interactions, could be
applicable to a cosmological model? What could be the main physical consequences of such a model?
In low-energy particle physics, there is a model of this type. This is the sigma-model \cite{ref10,ref11},
which invokes a spontaneously-broken chiral symmetry involving a scalar sigma, the pseudoscalar pion,
and the spin-1/2 nucleon fields. The pion is the massless Goldstone boson \cite{ref12} (in the absence of
quark masses), implied by the spontaneous symmetry breaking, i.~e.~by the non-zero vacuum expectation
value of the sigma field, which gives mass to the nucleon. The purpose of this paper is to present in a transparent
(partly heuristic) manner, the main possible physical results of a cosmological model with a spontaneously-broken
chiral symmetry \cite{ref13}. The fluctuations in the scalar field $\delta \phi$, are identified as massive
inflatons ($m_\phi\sim 5\times 10^{11}\GeV$ is estimated). There is the possibility that inflaton quanta can be
(meta) stable, and constitute a part of dark matter in the present epoch \cite{ref13}. The fluctuations in the pseudoscalar field 
$\delta b$, are initially massless Goldstone bosons; the homogeneous  
component of the field variable $b$ is zero if CP invariance holds.

We allow for non-zero values of the homogeneous component of $b$, with a metastable value $F_b\sim 5.5\eV$ acquired in the
early universe. This gives rise to a small vacuum energy density of absolute magnitude
$|\rho_\Lambda| \sim | -\lambda F_b^4| \sim (3\times 10^{-14})(5.5\times 10^{-9}\GeV)^4 \sim
2.7 \times 10^{-47}\GeV^4$, where the dimensionless parameter $\lambda\sim 3\times 10^{-14}$ is
determined empirically \cite{ref14} by the deviations from uniform matter density in the early
universe, which are assumed \cite{ref8,ref14} to arise from small variations in the inflaton
field. A non-zero value of $F_b$ means that CP invariance is spontaneously broken, here in a cosmological model.
What is the meaning of the small value of $F_b$? We find an answer to this question by considering an additional, hypothetical
coupling of the $b$ field to an ordinary neutrino. Two results then follow; these are intimately related to the small
$\rho_\Lambda$.
\begin{itemize}
\item[(1)] Mass for an ordinary neutrino is generated by $F_b$. We estimate a probable (largest)
value to be of the order of $0.055\eV$, i.~e.~$m_\nu \sim g_\nu F_b$ for $g_\nu\sim 0.01$. \footnote{
In the chiral model, radiative corrections to the tree-level potential which we have calculated \cite{ref13},
can establish a maximum in the inflaton effective potential. Inflation can occur during an (indeterminate) interval
of time before the field $\phi$ ``slips'' from this maximum, moving to a potential minimum. In the renormalization-group
calculations which give the potential maximum at $\phi\cong M_P$, we found a representative coupling strength of
$\phi$ (and $b$) to a neutrino-like, massive lepton $L$, which is present in the chiral model and receives its
mass from the vacuum expectation value of $\phi$, to be a (dimensionless) $g_L$ of the order of $0.01$. \cite{ref13}}
\item[(2)] An asymmetry between the number of antineutrinos and the number of neutrinos can be generated
in the early universe, in a non-equilibrium situation. This asymmetry exists at the time of electroweak
symmetry breaking; thus, it might be partly transformed into a baryon-antibaryon asymmetry.
\end{itemize}
We estimate that the antineutrino-neutrino asymmetry could have a magnitude as large as $\sim 10^{-9}$. We emphasize that it is CP noninvariance
which generates the small neutrino effective mass through the small value of $F_b$.\footnote{
The seemingly very small energy scale \cite{ref16} whose fourth power gives $\rho_\Lambda$, is $\lambda^{1/4} F_b \cong 0.002\eV$.
However, (largest) ordinary neutrino (presumably $\nu_\tau$) mass is empirically \cite{ref17} at a scale of at least,
about 25 times this number. We establish an explicit connection between these two small, energy (mass) scales, via $F_b$ in
the chiral model.
} At the same time, a scalar coupling of $b$ quanta to neutrinos
must arise. This coupling, together with the primary, pseudoscalar coupling of $b$ quanta to a neutrino-like, massive
lepton which is present in the primary chiral model (acquiring its mass from the non-zero vacuum expectation value
of $\phi$), allows naturally for a CP-violating, antineutrino-neutrino asymmetry to be generated in the early
universe. 

The (hermitian) Lagrangian density ${\cal L}$, with a (linear) chiral symmetry, involves the scalar field $\phi$, the 
pseudoscalar field $b$, and a neutral lepton field $L$. The squared ``bare'' mass parameter is $\mu^2<0$;
$g_L$ and $\lambda$ are dimensionless coupling parameters, $>0$. The (particle) symbols denote field
variables.
\ba
{\cal L} &=& \frac{i}{2}\left( \ol{L}\gamma_\mu (\partial^\mu L)-(\partial^\mu \ol{L})\gamma_\mu L\right) + \frac{1}{2}
\left( (\partial_\mu\phi)^2 + (\partial_\mu b)^2\right)\nn\\
& & - \frac{\mu^2}{2}\left(\phi^2+b^2\right) - g_L \ol{L}L\phi-ig_L\ol{L}\gamma_5 L b - \lambda\left(\phi^2+b^2\right)^2
\ea
($\partial^\mu$ denotes $\partial/\partial x^\mu$; $\gamma_5^\dagger = \gamma_5$; explicit indication of the space-time $(x)$
dependence of fields is suppressed.) The chiral transformations which leave ${\cal L}$ invariant are (for infinitesimal $\beta$)
\ba
L\to L - \frac{i\beta}{2}\gamma_5 L &,& \ol{L} = \ol{L} - \frac{i\beta}{2}\ol{L}\gamma_5\nn\\
\phi\to \phi-\beta b &,& b\to b + \beta b
\ea
As a consequence, there is a conserved axial-vector current, $\partial^\mu A_\mu=0$, with
\be
A_\mu = \frac{\delta {\cal L}}{\delta(\partial_\mu\beta)} = -\ol{L}\gamma_\mu\gamma_5 L + \left((\partial_\mu \phi)b
-(\partial_\mu b)\phi\right)
\ee
(where $\delta$ denotes an infinitesimal variation).

Consider the addition to ${\cal L}$ of a symmetry-breaking piece ${\cal L}' (\phi,b)$, given by
\be
{\cal L}' = c_\phi \phi.
\ee
Then $\partial^\mu A_\mu \neq 0$; instead the current is ``partially conserved'' \cite{ref10,ref11}.
\be
\partial^\mu A_\mu = -\frac{\delta {\cal L}}{\delta\beta} = c_\phi b
\ee
There is a non-zero matrix element defined by
\be
\sqrt{2p_0^b}\langle 0 | A_\mu(x)|b\rangle = i(p_\mu^b)e^{-i(p_\mu^b)x^\mu} \phi_c
\ee
Differentiating (6) and using (5) gives
\be
c_\phi = (p^b)_\mu (p^b)^\mu \phi_c = m_b^2\phi_c
\ee

We rewrite ${\cal L}(\phi,b)+{\cal L}'(\phi,b)$ in terms of field fluctuations, that is $\phi\to \phi_c+\delta \phi$,
$b\to \delta b$, using the notation for the quantum fields, $\delta\phi = \tilde{\phi}$, $\delta b=\tilde{b}$.
\renewcommand{\theequation}{8a}
\ba
\left({\cal L}(\tilde{\phi},\tilde{b})+{\cal L}'(\tilde{\phi},\tilde{b})\right) &=& \frac{i}{2}\left(
\ol{L}\gamma_\mu (\partial^\mu L)-(\partial^\mu \ol{L})\gamma_\mu L\right) - m_L \ol{L}L\nn\\
&+& \frac{1}{2}\left((\partial_\mu\tilde{\phi})^2 - m_\phi^2 \tilde{\phi}^2\right) + \frac{1}{2}
\left((\partial_\mu \tilde{b})^2 - m_b^2 \tilde{b}^2\right) \nn\\
&-& g_L \ol{L}L\tilde{\phi} - ig_L \ol{L}\gamma_5 L \tilde{b} - \lambda\left(\tilde{\phi}^2 + \tilde{b}^2\right)^2\nn\\
&-& 4 \lambda \phi_c \tilde{\phi}(\tilde{\phi}^2+\tilde{b}^2) + (c_\phi-m_b^2\phi_c)\tilde{\phi}
\ea
with masses given by
\renewcommand{\theequation}{8b}
\ba
m_L &=& g_L\phi_c\nn\\
m_\phi^2 &=& (\mu^2 + 12\lambda\phi_c^2)\nn\\
m_b^2 &=& (\mu^2 + 4 \lambda\phi_c^2)
\ea
In (8), $\tilde{\phi}$ and $\tilde{b}$ are not treated symmetrically, because $\phi_c \neq 0$. Setting the 
coefficient of the last term in (8a) to zero \cite{ref11}, gives $c_\phi=m_b^2\phi_c$,
as in (7). For the limiting situation \cite{ref11} in which ${\cal L}'$ in (4) vanishes, $c_\phi\to 0$.
For $\phi_c\neq 0$, $m_b^2=0$. This is the Goldstone mode for spontaneous symmetry breaking \cite{ref11},
which implies the existence of a massless pseudoscalar particle \cite{ref12}, described by the field $\tilde{b}$.
From (8b), we then have $\mu^2 = -4\lambda\phi_c^2$, and thus $m_\phi^2 = 8\lambda\phi_c^2 = -2\mu^2$
($ m_\phi\sim 4.9\times 10^{11}\GeV$ for $\phi_c\sim 10^{18}\GeV$ \cite{ref13}, and $\lambda\sim 3\times 10^{-14}$
\cite{ref14}). So $\phi_c^2 = -\mu^2/4\lambda$. The same result for $\phi_c$ is obtained by setting $b=0$
in (1), and then minimizing the effective potential-energy density, $(\frac{\mu^2}{2}\phi^2 + \lambda\phi^4)$.

We consider small non-zero values of the homogeneous field $b$, initially by adding to ${\cal L}$ in (1)
a small piece $-\frac{\mu_b^2}{2}b^2$ with $\mu_b^2<0$, and minimizing the small residual, effective potential-energy
density $(\frac{\mu_b^2}{2}b^2 + \lambda b^4)$. Parallel to the above result for $\phi_c$,
$ F_b^2 = \left|\langle 0|b|0\rangle \right|^2=-\mu^2_b/4\lambda$, with $m_b^2=8\lambda F_b^2 = -2\mu_b^2$.
The non-zero value of $F_b$ means that CP invariance is spontaneously broken in the early universe.
This is made manifest for the interactions by the unitary transformation \cite{ref15}
\setcounter{equation}{8} \renewcommand{\theequation}{\arabic{equation}}
\ba
L &\to& {\mathrm{e}}^{-i\gamma_5 \alpha/2} L \nn\\
\ol{L} & \to & \ol{L}{\mathrm{e}}^{-i\gamma_5\alpha/2}
\ea
with the choice $\tan\alpha=(g_L F_b)/m_L$. Then
\ba
&&\left\{ m_L\ol{L}L + g_L \ol{L}L\tilde{\phi} + i g_L \ol{L}\gamma_5 L \tilde{b}\right\}\nn\\
&\to& \left\{ M_L \ol{L}L + g_L(\cos\alpha \ol{L}L - i \sin\alpha \ol{L}\gamma_5 L )\tilde{\phi}\right.\\
&+& \left. g_L(i \cos\alpha \ol{L}\gamma_5 L + \sin\alpha \ol{L}L)\tilde{b}\right\}\nn
\ea
with $M_L = \sqrt{m_L^2 + (g_L F_b)^2}$, $\cos\alpha = m_L/M_L $, $\sin\alpha = (g_L F_b)/M_L$.
Clearly, the interactions of both $\tilde{\phi}$ and $\tilde{b}$ with $L$, violate CP invariance
when $F_b\neq 0$ (so $\alpha\neq 0$). This model has a small, residual \footnote{
Residual means after normalizing the negative, minimum potential-energy density, originating
in the $\phi_c^4$ term, to zero, with a positive, constant addition to $-{\cal L}$. For an
example where this not necessary because the minimum is calculated to be zero, see
the added note and the accompanying erratum figure for $V_c(\phi)$, in the first paper of Ref.~13
} vacuum energy density,
related to $F_b$, and to $\lambda$. The magnitude is
\be
|\rho_\Lambda|=\left| - \frac{|\mu_b^2|}{2}F_b^2 + \lambda F_b^4\right| = |-\lambda F_b^4|
\ee

A central purpose of this paper is to introduce a new dynamical element into the model, in an
attempt to set the small scale of $F_b$, independently of $|\rho_\Lambda|$. Consider
an ordinary neutrino (the heaviest), with ``bare'' mass $\tilde{m}_\nu$. Consider an additional,
small Lagrangian density
\be
-{\cal L}_\nu = \tilde{m}_\nu \ol{\nu}\nu + ig_\nu \ol{\nu}\gamma_5 \nu b
\ee
With the unitary transformation
\ba
\nu &\to& {\mathrm{e}}^{-i\gamma_5 \alpha_\nu/2}\nu\nn\\
\ol{\nu} &\to& \ol{\nu} {\mathrm{e}}^{-i\gamma_5 \alpha_\nu/2}
\ea
for $\tan\alpha_\nu=g_\nu F_b/\tilde{m}_\nu$, $-{\cal L}_\nu$ becomes
\be
-{\cal L}_\nu = m_\nu \ol{\nu}\nu + g_\nu (i\cos\alpha_\nu  \ol{\nu}\gamma_5\nu +
\sin\alpha_\nu \ol{\nu}\nu) \tilde{b}
\ee
with 
\ba 
&&m_\nu = \sqrt{\tilde{m}_\nu^2 + (g_\nu F_b)^2}\nn\\
&&\cos\alpha_\nu = \frac{\tilde{m}_\nu}{m_\nu}, \;\;\;\; \sin\alpha_\nu = \frac{g_\nu F_b}{m_\nu}\nn
\ea
Consider the situation if $\tilde{m}_\nu \ll (g_\nu F_b)$. Then, $\cos\alpha_\nu\sim 0$, $\sin\alpha_\nu
\sim 1$. The effective mass of the neutrino is then $\sim (g_\nu F_b)$. There is a scalar
interaction with the pseudoscalar $\tilde{b}$, which when acting together with the (dominant) interaction
$g_L i  \cos\alpha \ol{L}\gamma_5 L \tilde{b}$ in (10), violates CP invariance. An empirical \cite{ref17},
(largest) ordinary neutrino ( presumably $\nu_\tau$) mass may be $\sim 0.055\eV$.\footnote{
Assuming a hierarchal structure for ordinary neutrino masses, as for quark masses. The scale of neutrino
mass could reach up to some eV, since present data concerns differences in squared masses, the largest
of which is $\sim 3\times 10^{-3} \eV^2$. \cite{ref17,ref18}
} With
$g_\nu \sim 0.01$ $^{F1}$, this implies $F_b\sim 5.5 \eV$. Using this value for $F_b$ in (11), with
an empirical \cite{ref14} $\lambda\sim 3\times 10^{-14}$, yields
\be
|\rho_\Lambda| = |-\lambda F_b|^4 \sim 2.7\times 10^{-47} \GeV^4
\ee
This is about 0.7 of critical density. Clearly, the very small value of $\lambda$ plays an essential
roll in this numerical estimate of $|\rho_\Lambda|$. Perhaps the most direct and simple way to
ensure that this small $\rho_\Lambda$ is $>0$, is to explicitly break the symmetry by allowing $\lambda$
to go to $-\lambda$ in only the term $-\lambda b^4$ in ${\cal L}$ in (1). 
Then $\mu_b^2$ is $>0$; it can be interpreted as the squared mass of the $b$ quanta. $F_b$ must be viewed
as a metastable maximum \footnote{
A necessary shift of the second derivative of the effective potential, $(\mu_b^2-12\lambda F_b^2)=-2\mu_b^2$
to zero, could be brought about by a mixing of $b$ with another (initially massless) pseudoscalar, at
a time $t_b\sim 1/m_b \sim 10^{-10}\secs$ (the universe time is about the inverse of the Hubble expansion
rate $H(t)$; so $H(t_b)$ has fallen to $\sim m_b$). This time is approaching the time associated with the quark-hadron
transition, with appearance of a Goldstone pion. Such mixing would raise the physical squared mass of $b$ (
from $\mu_b^2$ to about $3\mu_b^2$ is required).
}
; some decrease with time of $\rho_\Lambda$, and of $m_\nu$, is then possible.
This is a problematic aspect of this type of  model, with negative, squared bare-mass parameters.$^{F3}$
However, the above symmetry breaking is ameliorated by the fact that there are two spin-zero fields at different
energy scales. A dominant, positive-definite energy density clearly resides in the $\lambda\phi^4$  term
for large field values at the earliest times ($\phi\sim M_P$). We have achieved a definite, conceptual connection
between possible empirical energy scales for ordinary neutrino mass\cite{ref17}, and for $\rho_\Lambda$ \cite{ref6,ref7,ref14}.
The Goldstone pseudoscalar can have mass of the order of $2\sqrt{2}\sqrt{\lambda}F_b \sim 2.7\times 10^{-6}\eV$.
Achieved in addition, via the explicit interactions that we have delineated above, is the natural possibility
of an antineutrino-neutrino symmetry having been created in the early universe. Before we make a definite estimate,
we note that although $F_b\sim 5.5 \eV$ is seemingly a very small energy (mass) scale,\footnote{
It is worth noting that a time scale $1/F_b \sim 1.2\times 10^{-16}\secs$ is near to the time of electroweak symmetry
breaking. Also note that a mass of order $(F_b/2m_K)F_b \sim 3\times 10^{-8}\eV$, is similar to the mass scale
which characterizes $K_L^0-K_S^0$ mixing. An intermediate $b$ could contribute to this mixing. }
as a scale for cosmological
CP violation, it is a much larger scale than a mass scale which characterizes the empirical, indirect CP violation
\cite{ref17} in the mixing of the $K_L^0-K_S^0$ states, i.~e.~$\sim \sqrt{2}|\epsilon| \Gamma_{K_S^0}/2 \sim 1.2 \times 10^{-8}\eV$
(the decay half-width $\Gamma_{K_S^0}/2 \cong 3.7\times 10^{-6}\eV$; the CP-violating mixing parameter $|\epsilon| \cong 2.3\times 10^{-3}$).
The hypothetical, specific primary mechanism for an asymmetry that we consider is illustrated by the diagrams in Figs.~(1a,b).
Fig.~(1a) shows the tree amplitude for decay of a massive scalar $S$ to $L+\ol{\nu}$, or to $\ol{L}+\nu$.
Fig.~(1b) shows a final-state interaction (radiative correction) in this amplitude, due to exchange of a $b$ quantum,
using at the lower vertex, the scalar coupling to $\ol{\nu}(\nu)$ in (14), and at the upper vertex, the pseudoscalar coupling
to $L(\ol{L})$ in (10). A real intermediate state in the final-state interaction gives rise to a factor of $i$,
which together with the CP violation from the $b$-exchange vertices, results in an interference term in the absolute
square of the amplitude from Figs.~(1a,b) which differs in sign for $S\to L+\ol{\nu}$ and for $S\to \ol{L}+\nu$.
This naturally generates an asymmetry between the number of antineutrinos and neutrinos. The hermitian, CP-invariant
($B$ is real), decay interaction is written as
\be
g_S\left\{ \ol{L}\frac{(1-B\gamma_5)}{2}\nu + \ol{\nu}\frac{(1+B\gamma_5)}{2}L\right\}S
\ee
The decay partial width is estimated as 
\be
\Gamma(L\ol{\nu}+\ol{L}\nu) \sim 10^{-2}g_S^2 m_L \sim \frac{1}{10^{-36} \secs}
\ee
using representative mass values at an energy scale of $10^{16}\GeV$:\footnote{
For $\phi_c\sim 10^{18}\GeV$ and $g_L\sim 0.01$ \cite{ref13}, we have $m_L\sim 10^{16}\GeV$ from (8b).
}  $m_L\sim 10^{16}\GeV$, 
$m_\nu\cong 0$, $m_S\sim 1.4\times 10^{16} \GeV$, and $g_S^2 \sim 0.005$. We assume that, although there are
other decay modes of $S$, these do not involve the explicit $\ol{\nu}$-$\nu$ asymmetry of the
two primary modes estimated here (for example, like $\gamma\gamma$, $ZZ$, $W^-W^+$), and that the subsequent disappearance
of $L$, $\ol{L}$ into radiation, does not vitiate the explicit primary, $\ol{\nu}$-$\nu$ asymmetry.
If the branching ratio $(\mathrm{(b.r.)}_{S\to\ol{\nu},\nu}$, for the two modes estimated above is of order $10^{-1}$, then
the lifetime $\tau_S$ is of order $10^{-37}\secs$ (this is an ``initial'' time interval after hypothetical inflation).
The Hubble expansion time scale corresponding to an energy scale of $\sim 10^{16}\GeV$ is $\sim 10^{-38}\secs$. The $S$ decay
thus occurs in a non-equilibrium situation; back-reactions do not vitiate the primary $\ol{\nu}$-$\nu$ asymmetry.
From an estimate of the absorptive part of the diagram in Fig.~(1b), we calculate this number asymmetry in $N_{\ol{\nu},\nu}$
(for one type of neutrino), relative to photon number $N_\gamma$,
\ba
|{\cal A}_\nu| &=& \frac{\left| N_{\ol{\nu}}-N_\nu\right|}{N_\gamma} \sim \left| B\left(\frac{g_L g_\nu}{18\pi}\right)
\right| (\mathrm{(b.r.)}_{S\to \ol{\nu},\nu} \left(\frac{N_S}{N_\gamma}\right)\nn\\
&\sim& \left( 2\times 10^{-6}\right)\left(10^{-1}\right)\left(\frac{N_S}{N_\gamma}\right)\sim 2\times 10^{-9}
\ea
The first factor in (18) arises explicitly from the CP-violating interference between the amplitudes from Figs.~(1a,b).
The numerical estimate of this factor is for $|B|\sim 1$, $|g_L g_S|\sim (10^{-2})^2 \sim 10^{-4}$, $^{F1}$
$p_\nu\sim 0.35\times 10^{16}\GeV$ ( $m_S = p_\nu + \sqrt{p_\nu^2+m_L^2}$ ), and $m_b \sim 2.7 \times 10^{-6}\eV$ as
estimated above. Two aspects of this factor are noteworthy
\begin{itemize}
\item[(1)] There is no enhancement because of the long range of the final-state interaction
from exchange of a very low-mass $b$ quantum \footnote{
Similarly for the high-energy interaction between at least the heaviest of ordinary neutrinos (presumably $\nu_\tau$), due
to $b$ exchange. There is a long-range interaction between cosmic-background $\nu_\tau$ with momenta much less than their mass.
}.
\item[(2)] The small energy scale $F_b$ does not appear explicitly. This is because it does not appear in the scalar
interaction term in (14), for $\tilde{m}_\nu \ll (g_\nu F_b)$ (i.~e.~$\sin\alpha_\nu \cong 1$). 
\end{itemize}
The last factor in 
(18) is $(N_S/N_\gamma)$, the ratio of the original number of decaying $S$ to the number of (eventual) photons 
$(\sim 10^{88})$. We have used $\sim 10^{-2}$ for this ratio, because this corresponds to an energy density
of $\sim 10^{62}\GeV^4$ in $S$ particles at $\sim 10^{-38}\secs$, near to their decay time. This energy density is
comparable to the potential-energy density in the inflaton field at its ``initial'' position $\phi\sim M_P$, 
which is a calculated \cite{ref13} potential maximum: $V(\phi\sim M_P)\sim \lambda M_P^4 \sim 6\times 10^{62}\GeV^4$
(for $\lambda\sim 3\times 10^{-14}$). This opens a possibility  that $S$ quanta are produced by a coupling to the energy
of motion, as the inflaton field ``falls'' into the potential minimum at $\phi_c\sim 10^{18}\GeV$. \cite{ref13} The
primary energy density in radiation can thus originate from decays of these $S$ quanta. 

In summary, the magnitude of the asymmetry estimated in (18), would seem to allow for the possibility
that a part of this $\ol{\nu}$-$\nu$ asymmetry is transformed into a baryon-antibaryon asymmetry, by hypothetical
processes \cite{ref19} which take place near to the time of electroweak symmetry breaking, $\sim 10^{-12}\secs$.

Before concluding, we comment briefly on the possibility that inflaton quanta with mass of $\sim 5\times 10^{11}\GeV$,
could have a lifetime much larger than the present age of the universe, and so could constitute a part of dark
matter today \cite{ref13}. If inflaton decay were to occur only indirectly via a triangle diagram involving
$\phi$ coupling to $L\ol{L}$, followed by $S$ exchange, resulting in a $\nu\ol{\nu}$ final state \cite{ref13},
then there would be a natural reason for suppression of the matrix element. The effective quantum interaction
would have the structure $\ol{\nu}(1+B\gamma_5)(1-B\gamma_5)\nu\tilde{\phi}$, which vanishes for $|B|=1$. However,
in the present model with a nearly massless $b$, $\phi$ decays rapidly to $bb$ pairs, via the trilinear interaction
in the next-to-last term in  (8a). Only if this term were to be removed \footnote{
This involves a degree of ``fine-tuning'', which is however ameliorated by the smallness of $\lambda$.
}, perhaps by explicit symmetry breaking
that also removes tadpole diagrams \cite{ref11} which arise from the trilinear bosonic couplings, could the inflaton
be (meta)stable (against sole decay into $\nu\ol{\nu}$). In any case, there is an upcoming, direct experimental \cite{ref20}
test of the idea that dark matter decays very slowly into neutrino-antineutrino pairs \cite{ref13}. Neutrinos with 
energies just above $10^{20} \eV$, should be detectable as initiators of the highest-energy cosmic-ray air showers \cite{ref21},
either in near-horizontal showers, or in more vertical showers, if neutrinos at this energy scale have anomalous, fairly
strong interactions \cite{ref22}. High energy ($>10^{16}\eV$) air showers also offer the unique possibility to look for
arriving Goldstone bosons which might have fairly strong interactions in air, with unusual dynamical characteristics \cite{ref23}.

To summarize, we have sought a model that naturally contains two spin-zero fields, which could play a role in the evolution
of the universe. In a cosmological model with an initial chiral symmetry, there are two dynamically-related fields
with spin zero. We have identified the scalar field with the inflaton \cite{ref13}. We have examined the possible
dynamical role of a pseudoscalar Goldstone field which arises as a consequence of spontaneous symmetry breaking.
The homogeneous component of the pseudoscalar field could provide a residual, small vacuum energy density, whose
early magnitude is similar to the energy density presently attributed to an  effective cosmological constant.
The new idea which we have introduced here is the separate determination of the small, vacuum energy scale $F_b$,
by relating this scale of a few eV dynamically to the empirical, small mass scale of ordinary neutrinos.
The model involves a (cosmological) spontaneous breakdown of CP invariance, through $F_b\neq 0$. This allows
naturally for the generation of an antineutrino-neutrino asymmetry in the early universe, whose magnitude we have
estimated to be significant for a possible subsequent, partial transformation into a baryon-antibaryon asymmetry.
The small asymmetry, the small effective neutrino mass, and the small, residual vacuum energy are connected to the
small, cosmological breakdown of CP invariance. In the model, these quantities would all be zero if CP invariance held.

The ratio $r_{CP}=F_b/\phi_c \sim F_b/M_P \sim 4.6 \times 10^{-28}$ is set by empirical neutrino mass (for the scale
of $F_b$), and by inflaton mass (for the scale of $\phi_c$, given  $\lambda$). Raising $r_{CP}$ to the fourth
power yields $\sim 4.5\times 10^{-110}$, which is a dimensionless number identical to a frequently mentioned \cite{ref3},
miniscule number that characterizes one form of the ``cosmological-constant problem''. This is the dimensionless
discrepancy factor between the presently ``observed'' small value of the vacuum energy density, $\sim 3\times
10^{-47}\GeV^4$, and a possible ``a priori'' value, $\sim \lambda M_P^4\sim 6\times 10^{62}\GeV^4$. From the point
of view of this paper, the latter energy density is at the origin of early inflation and the primary radiation
energy density, whereas the former, residual energy density has a different, but related (through $\lambda$)
origin, and appears as an effective cosmological constant.

\newpage
\begin{figure}[t]
\begin{center}
\mbox{\epsfysize 13cm \epsffile{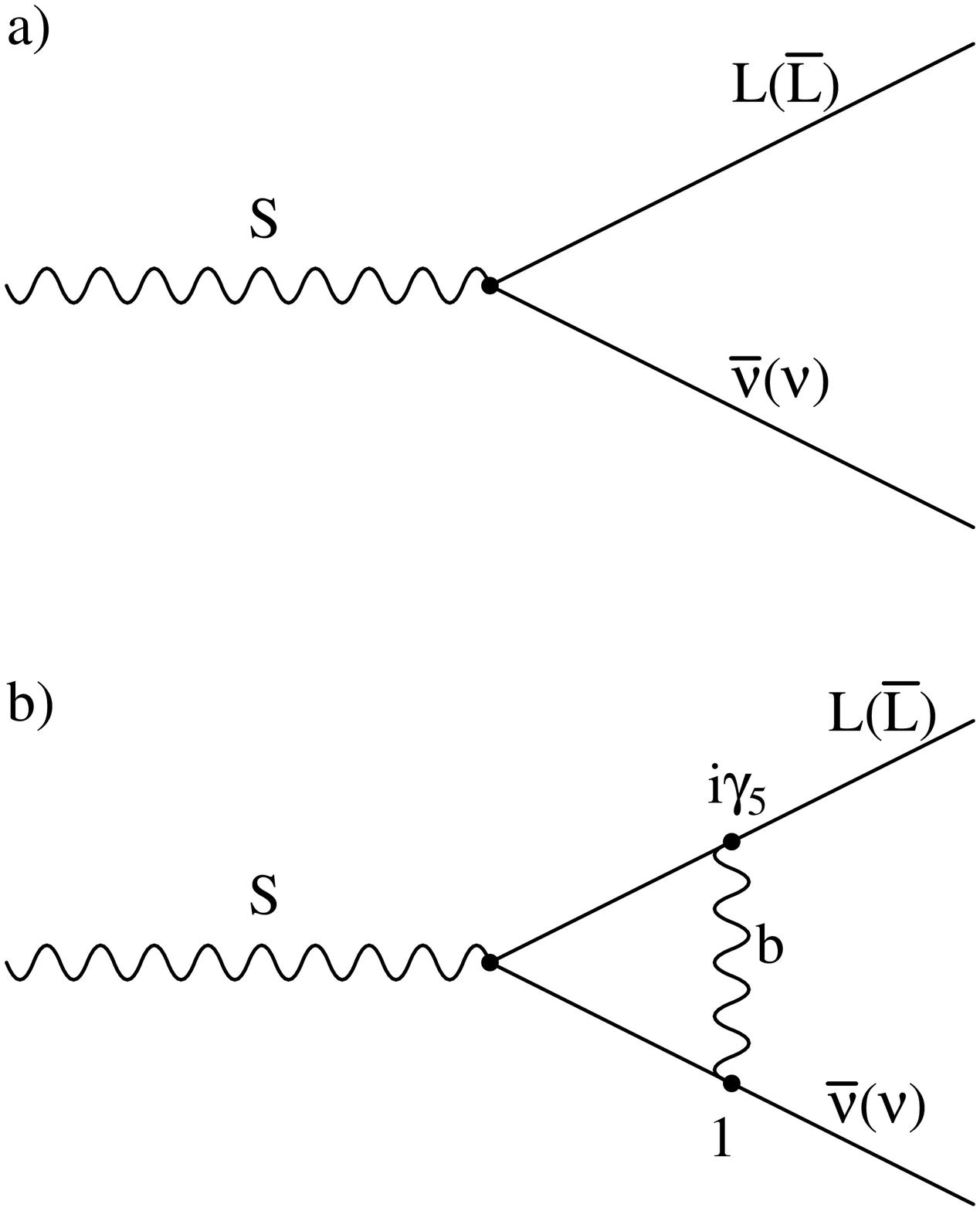}}
\end{center}
\caption{a) The tree-level diagram for the specific decay of $S\to L+\ol{\nu}\;(\ol{L}+\nu)$. b) The additional diagram
involving a final-state interaction in the real intermediate state of $L+\ol{\nu}\;(\ol{L}+\nu)$, via $b$ exchange.
The CP noninvariance arises explicitly from appearance of $\i\gamma_5$ at the upper vertex, and 1 at
the lower vertex. The interference between the two diagrams is opposite in sign for $L\ol{\nu}$ and
for $\ol{L}\nu$. This gives rise to a $\ol{\nu}$-$\nu$ asymmetry.}
\end{figure}

\begin{thebibliography}{9}
\bibitem{ref1} P.~Jordan, {\it Schwerkraft und Weltall}, Friedrich Vieweg und Sohn, Braunschweig, 1955;\\
Z.~Phys.~{\bf 157}, 112 (1959)
\bibitem{ref2} C.~Brans and R.~H.~Dicke, Phys.~Rev.~{\bf 126}, 925 (1961)
\bibitem{ref3} B.~Ratra and P.~J.~E.~Peebles,  Phys.~Rev.~{\bf D37}, 3406 (1988)
\bibitem{ref4} G.~W.~Anderson and S.~M.~Carroll, astro-ph/9711288v2\\
T.~Damour, G.~W.~Gibbons, and C.~Grundlach, Phys.~Rev.~Lett.~{\bf 64}, 123 (1990)\\
J.-P.~Uzan, Rev.~Mod.~Phys.~{\bf 75}, 403 (2003) and references therin
\bibitem{ref5} A.~Linde, {\it Particle Physics and Inflationary Cosmology}, Harwood Acadamic Publishers GmbH,
CH-7000 Chur, Switzerland 1990
\bibitem{ref6} S.~Perlmutter et al., Nature (London) {\bf 391}, 51 (1998)
\bibitem{ref7} A.~G.~Riess et al., Astron.~J.~{\bf 116}, 1009 (1998)
\bibitem{ref8} P.~J.~E.~Peebles and A.~Vilenkin, Phys.~Rev.~{\bf D59}, 063505 (1999)
\bibitem{ref9} G.~R.~Farrar and P.~J.~E.~Peebles, astro-ph/0307316v2, Sept.~2003
\bibitem{ref10} M.~Gell-Mann and M.~Levy, Nuovo Cimento {\bf 16}, 705 (1960)
\bibitem{ref11} B.~W.~Lee, {\it Chiral Dynamics}, Gordon and Breach Science Publishers, Inc., 1972
\bibitem{ref12} J.~Goldstone, Nuovo Cimento {\bf 19}, 155 (1961)
\bibitem{ref13} S.~Barshay and G.~Kreyerhoff, Z.~Phys.~{\bf C75}, 165 (1997); Erratum Z.~Phys.~{\bf C76}, 577 (1997)\\
S.~Barshay and G.~Kreyerhoff, Eur.~Phys.~J.~{\bf C5}, 360 (1998)
\bibitem{ref14} H.~Kurki-Suonio and G.~J.~Mathews, Phys.~Rev.~{\bf D50}, 5431 (1994)
\bibitem{ref15} T.~D.~Lee, Phys.~Rep.~{\bf 9C}, 143 (1974)\\
T.~D.~Lee, {\it Particle Physics and Introduction to Field Theory}, Harwood Academic Publishers GmbH,
CH-7000 Chur, Switzerland 1981, chapter 16
\bibitem{ref16} P.~J.~E.~Peebles, Astrophys.~J.~{\bf 284}, 439 (1984)
\bibitem{ref17} Particle Data Group, {\it Review of Particle Physics}, Phys.~Rev.~{\bf D66}, 010001 (2002)
\bibitem{ref18} S.~M.~Bilenky, G.~Giunti, J.~A.~Grifols and E.~Masso, hep-ph/0211462, Nov.~2002
\bibitem{ref19} G.~'t Hooft, Phys.~Rev.~Lett.~{\bf 37}, 8 (1976);
Phys.~Rev.~{\bf D14} 3432 (1976)\\
V.~A.~Kuzmin, V.~A.~Rubakov and M.~E.~Shaposhnikov, Phys.~Lett.~{\bf B155}, 36 (1985)\\
V.~A.~Rubakov, Nucl.~Phys.~{\bf B256}, 509 (1985)
\bibitem{ref20} M.~Nagano and A.~A.~Watson, Rev.~Mod.~Phys.~{\bf 72}, 689 (2000)
\bibitem{ref21} S.~Barshay and G.~Kreyerhoff, Nuovo Cimento {\bf A112}, 1463 (1999)
\bibitem{ref22} S.~Barshay and G.~Kreyerhoff, Phys.~Lett.~{\bf B535}, 201 (2002)
\bibitem{ref23} S.~Barshay and G.~Kreyerhoff, Mod.~Phys.~Lett.~{\bf A16}, 1061 (2001). Possible
unusual dynamics in air showers initiated by a hypothetical Goldstone boson associated
with an energy scale $F$ as low as a few eV, is discussed in this paper in connection
with certain experimental results.
\end{thebibliography}
\end{document}